\documentclass[%
twocolumn,
superscriptaddress,
notitlepage,
showpacs,
amsmath,amssymb,
aps,
prb,
floatfix,
longbibliography
]{revtex4-2}

\usepackage{graphicx}
\usepackage{xcolor}
\usepackage{hyperref}
\usepackage{enumitem}

\usepackage{braket}
\usepackage[thinc]{esdiff}

\begin{document}

	\title{Particle pairing causes subdiffusion of heavy particles in the imbalanced Hubbard model}
	\author{Mirko Daumann}\email{mdaumann@physik.uni-bielefeld.de}
	\affiliation{
		Fakult\"at f\"ur Physik, Universit\"at Bielefeld, Postfach 100131, D-33501 Bielefeld, Germany
	}
	\author{Thomas Dahm}
	\affiliation{
		Fakult\"at f\"ur Physik, Universit\"at Bielefeld, Postfach 100131, D-33501 Bielefeld, Germany
	}
	
	\date{\today}
	
	\begin{abstract}
	  The imbalanced Hubbard model features a transition between dynamic regimes depending on the mass ratio and coupling strength between two different particle species.
	  A slowdown of the lighter particle transport can be attributed to an emergent effective disorder induced by the heavy particles for high mass ratio and strong coupling.
	  This subdiffusive regime has been interpreted as a Griffiths phase, linking the effect to the coexistence of metallic and insulating regions.
	  Here, we investigate the dynamics of the heavy particles, which also reveals subdiffusive behavior, yet cannot be explained within the Griffiths picture.
	  We demonstrate that heavy particles predominantly form small clusters, mainly pairs, during the dynamical process, which reduces their propagation speed and transiently shifts the time-dependent diffusion constant into the subdiffusive regime at late times.
	  The necessary attraction between particles driving this process can be understood within the Born-Oppenheimer approximation.
	  We introduce a classical one-dimensional random walk model that can quantitatively reproduce the subdiffusion dynamics in the strong coupling regime. 
	\end{abstract}
	\maketitle
	The imbalanced Hubbard model serves as an example of a system that exhibits a transition from regular diffusion to localization without external quenched disorder~\cite{2014Grover,2015Jin,2019Sirker,2020Heitmann,2022Oppong,2022Zechmann,2024Kiely}.
	It consists of a mixture of light and heavy particles coupled via repulsive on-site interaction.
	At sufficiently strong mass imbalance and interaction strength, light particles begin to localize in a quasi-static disorder potential induced by the heavy particles, thereby slowing down their transport.
	This subdiffusive regime for light particles has been interpreted as a Griffiths phase, which provides an intuitive explanation in terms of coexisting metallic and insulating regions~\cite{2010Vojta,2015Agarwal}.
	
	However, the dynamic properties of the heavy particle species have received significantly less attention and are the main focus of this work.
	We first show that heavy particles form small clusters, particularly pairs.
	This clustering effect arises from the repulsive interaction, which induces an effective attractive Casimir-like force between heavy particles due to the response of the light subsystem when the coupling strength exceeds the kinetic energy of the light particles.
	Such stable clusters have been observed previously~\cite{2006Winkler,2012Deuchert,2024Wang} and are interpreted here within the Born-Oppenheimer approximation~\cite{1927Born,2024Daumann}.
	
	The idea that cluster formation processes slow down transport has been proposed for systems with long-ranged interactions~\cite{1984Kagan,2023Li}.
	We show that a similar effect arises naturally for the short-ranged interaction present in the imbalanced Hubbard model.
	We demonstrate that the pairing regime, defined by the interaction strength, coincides with transient subdiffusive transport - a phenomenon not captured by the Griffiths picture.
	This interpretation is further supported by a random walk model, where a clustering mechanism is explicitly implemented and quantitatively reproduces the observed subdiffusive behavior.
	
\begin{figure}
		\includegraphics[width=1.0\linewidth]{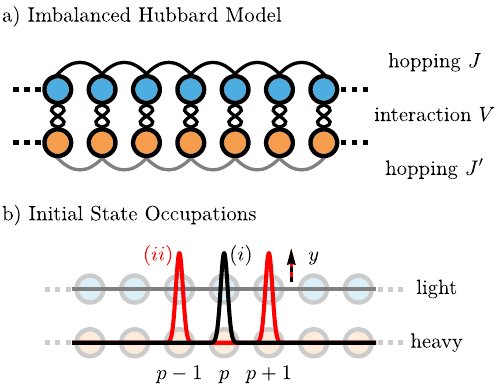}
		\caption{a) Illustration of the imbalanced Hubbard model, showing hopping and interaction strengths. b) Particle densities of initial states with (i) a single peak at position $p$ and (ii) double peaks at $p-1$ and $p+1$. The parameter $y$ induces an out-of-equilibrium density shift at the respective orbitals. All other sites are filled with random background occupations, fluctuating only slightly around a constant value $n_\text{bg}$ due to the sufficiently large Hilbert space dimension.}
		\label{fig:model}
	\end{figure}

	The imbalanced Hubbard Hamiltonian (Fig.~\ref{fig:model}a) is given by
	\begin{equation}
		\begin{split}
			\mathcal{H}=\sum_{i=1}^{L}\Big(&-J\cdot\hat{\mathrm{c}}^\dagger_{i}\hat{\mathrm{c}}^{}_{i+1}-J'\cdot\hat{\mathrm{d}}^\dagger_{i}\hat{\mathrm{d}}^{}_{i+1}\\
			&+V\cdot\hat{\mathrm{c}}^\dagger_{i}\hat{\mathrm{c}}^{}_{i}\hat{\mathrm{n}}^{}_{i}\Big)+\text{h.c.}
		\end{split}
		\label{eq:hamiltonian}
	\end{equation}
	with $\hat{\mathrm{c}}^\dagger_{i}$ ($\hat{\mathrm{c}}^{}_{i}$) denoting fermionic creation (annihilation) operators for the light, and $\hat{\mathrm{d}}^\dagger_{i}$ ($\hat{\mathrm{d}}^{}_{i}$) for the heavy particle species. 
	The occupation number operator is defined for heavy particles as $\hat{\mathrm{n}}^{}_{i}=\hat{\mathrm{d}}^\dagger_{i}\hat{\mathrm{d}}^{}_{i}$.
	The hopping amplitude of light particles is set to $J = 1$ as the unit of energy, while the kinetic energy of heavy particles is given by $J' \ll J$.  
	Both species are coupled through a repulsive on-site interaction of strength $V$. 
	We focus on the variation of $V$, while fixing $J'=0.01$.  
	The lattice consists of $2L$ orbitals with periodic boundary conditions and is occupied by up to $L$ particles.
	The Hilbert space dimension is given by $d = \sum_{m=0}^{L} \binom{2L}{m}$.
	$L$ is set to $12$ which yields $d \approx 10^7$.
	The resulting filling fraction, in general, varies depending on the choice of initial states and system parameters.
	In the present cases, it averages between $45\%$ and $50\%$.
	
	Initial states for time evolution can generally be constructed as a superposition of equilibrium and out-of-equilibrium contributions~\cite{2018Reimann}:
	\begin{equation}
		\begin{split}
		\ket{\Psi}&=\ket{\Psi_\text{eq}}+y\cdot\ket{\Psi_\text{neq}}\\
		&=\left(e^{-\beta\mathcal{H}}+y\cdot e^{-\frac{\beta\mathcal{H}}{2}}\mathcal{A}_i\, e^{-\frac{\beta\mathcal{H}}{2}}\right)\ket{\Psi_\text{r}}\ .
		\end{split}
		\label{eq:initialStates}
	\end{equation} 
	Setting $y \neq 0$ drives the system out of canonical equilibrium according to the operator $\mathcal{A}_i$.
	These initial states satisfy the conditions of dynamical typicality~\cite{2009Bartsch}, where $\ket{\Psi_\text{r}}$ denotes a normalized complex random state drawn from a Gaussian distribution with mean 0.0 and standard deviation 1.0 in the particle number basis~\cite{2015Jin,2020Heitmann}.
	The $\ket{\Psi}$ states are normalized after summing the equilibrium and out-of-equilibrium contributions.
	Satisfying dynamical typicality ensures that their time evolution is representative of a large ensemble of pure states with similar initial characteristics.
	For each set of parameters, the data shown result from averaging the time evolutions of up to ten independently sampled $\ket{\Psi}$ states, which show only minor deviations from one another.
	Unless stated otherwise, we fix $y = 1$ as a constant, positioned at the upper end of the linear response regime, and consider the high-temperature limit $\beta = 0$.
	There are two different initial state settings:
		\begin{itemize}[noitemsep,topsep=0pt]
			\item[(i)] For the calculation of time-dependent diffusion constants $\mathcal{A}_1=\hat{\mathrm{n}}_{p}$ is used as out-of-equilibrium operator which creates a heavy particle occupation density peak at rung $p=\frac{L}{2}$.
			\item[(ii)] To study the clustering of heavy particles a double density peak formed by $\mathcal{A}_2=\hat{\mathrm{n}}_{p-1}+\hat{\mathrm{n}}_{p+1}$ is used (Fig.~\ref{fig:model}b).
		\end{itemize}
	The matrix exponentials for time evolution are solved numerically using Krylov subspace methods~\cite{1950Lanczos,1986Park,2003Moler,2006Mohankumar}.
	
	We investigate the time-dependent diffusion constant $D(t)$, which serves as an indicator for different dynamical phases.
	It is calculated as the time derivative of the square root of the mean squared displacement $\Sigma(t) = \sqrt{\braket{\hat{\mathrm{r}}^2(t)} - \braket{\hat{\mathrm{r}}(t)}^2}$ of heavy particles:
	\begin{equation}
		\begin{gathered}
			D(t)=\diff{}{t}\Sigma(t)\quad \text{with}\\
			\Sigma(t)=\sqrt{\sum_{i=1}^Li^2\cdot n_i(t)-\left(\sum_{i=1}^Li\cdot n_i(t)\right)^2}\ .
		\end{gathered}
		\label{eq:diffusionConstant}
	\end{equation}
	$n_i(t)\sim\braket{\hat{\mathrm{n}}_i(t)}-n_\text{bg}$ contains subtraction of random background occupations $n_\text{bg}$ to simulate single particle tracking, and is normalized: $\sum_in_i(t)=1$.
	$n_\text{bg}$ depends on the choice of $\mathcal{A}_i$ and $y$, and lies on average between $0.399$ and $0.422$.
	A diffusion constant of $D(t) = 0.5$ indicates regular diffusion, while $D(t) < 0.5$ and $D(t) > 0.5$ are characteristic of sub- and superdiffusion, respectively~\cite{2000Metzler,2017Lev,2019Oliveira}.
	
	In order to extract information on the degree of particle pairing and cluster formation, the following set of operators is defined:
	\begin{equation}
		\mathcal{C}_m=\sum_{i=1}^L(1-\hat{\mathrm{n}}_{i-1})(1-\hat{\mathrm{n}}_{i+m})\prod_{j=0}^{m-1}\hat{\mathrm{n}}_{i+j}
	\end{equation}  
	for $\ m\in\{1,\dots,L-1\}$.
	They count the number of isolated $m$-clusters in a basis state; for example, $\mathcal{C}_1$ gives the number of single isolated particles, $\mathcal{C}_2$ counts pairs, $\mathcal{C}_3$ counts triplets, and so on. 
	Periodic boundary conditions are applied to account for clusters crossing the boundary.
	The expectation values $c_m(t)$ are calculated using the heavy subsystem reduced density matrix $\rho_\text{H}$:
	\begin{equation}
		c_m(t)=\operatorname{Tr}\left(\rho_\text{H}(t)\cdot\mathcal{C}_m\right)
	\end{equation}
	where $\rho_\text{H}(t) = \operatorname{Tr}_\text{L} \left( \ket{\Psi(t)} \bra{\Psi(t)} \right)$ is obtained by tracing out the light particle subsystem.
	We furthermore consider the ratio $c_{n,m}(t)$ of clusters over time and the correlations $\rho_{n,m}$:
	\begin{equation}
			c_{n,m}(t)=\frac{c_n(t)-c_n(0)}{c_m(t)-c_m(0)}\ , \
		 	\rho_{n,m}=\frac{\operatorname{cov}_t\left(c_n(t),c_m(t)\right)}{\sigma_t(c_n(t))\cdot\sigma_t(c_m(t))}
	\end{equation}
	where $\operatorname{cov}_t$ denotes the covariance and $\sigma_t$ the standard deviations, both calculated for the respective cluster size time data.
	\begin{figure}
		\includegraphics[width=1.0\linewidth]{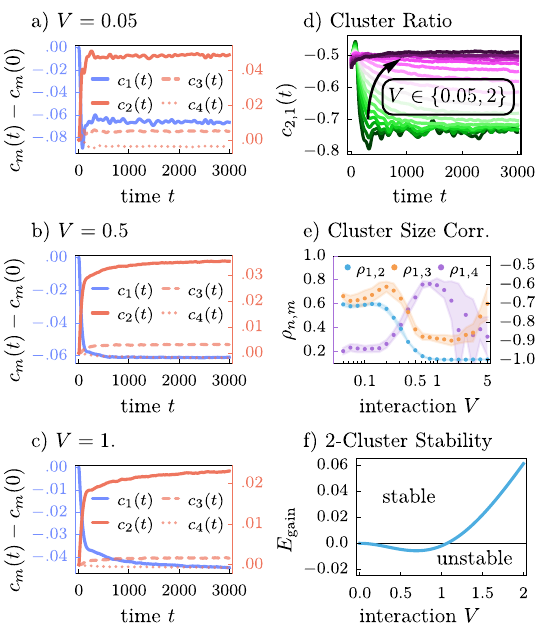}
		\caption{a-c) Cluster size expectation values $c_m(t)$ for different coupling strengths $V$. A significant anti-correlation between $c_1(t)$ and $c_2(t)$ becomes apparent for higher $V$ values. d) Cluster ratio $c_{n,m}(t)$ between 1- and 2-clusters. The figure shows 17 curves, with $V$ values increasing logarithmically from 0.05 (green) to 2 (pink). The arrow points in the direction of increasing $V$. Single particles forming pairs result in a constant $c_{2,1}(t) = -1/2$, which is fulfilled for higher $V$ (dark pink curves). e) Correlations $\rho_{n,m}$ between single particles and pairs, triplets (right scale), and quadruplets (left scale). The standard deviation from different random initial states is shown as bands of uncertainty. f) System's energy gain $E_\text{gain}$ from pairing, calculated in the Born-Oppenheimer approximation.}
		\label{fig:pairing}
	\end{figure} 
        
	{\it Evidence for Particle Pairing} - We begin by discussing the clustering of heavy particles by examining the time evolution of two initially separated density peaks (setting (ii)) while setting subdiffusion aside for now.
	The corresponding initial states are shown in Fig.~\ref{fig:model}b.  
	Cluster size data $c_m(t)$ for varying coupling strengths $V$ is presented in Figs.~\ref{fig:pairing}a–c.  
	At early times, the number of single particles $c_1(t)$ decreases sharply, while the number of pairs $c_2(t)$ increases correspondingly with approximately half the amplitude.
	As the coupling increases, these processes slow down while becoming noticeably more correlated.
	The amplitudes of triplets and quadruplets are less pronounced and flatten further with increasing coupling.
	Among the observed processes, the pairing of single, isolated particles, $c_1(t) \rightarrow c_2(t)$, is the most prominent.
	This is particularly evident in Fig.~\ref{fig:pairing}d, which shows the time-dependent ratio $c_{n,m}(t)$ of paired to single particles.  
	A perfect transition, where two single particles form one pair, corresponds to a constant ratio of $c_{2,1}(t) = -\frac{1}{2}$, which aligns increasingly well with the data as the coupling strength grows.
	The correlations $\rho_{n,m}$ between different cluster sizes, shown in Fig.~\ref{fig:pairing}e, support the previous observations.  
	In particular, $\rho_{1,2}$ exhibits a perfect anti-correlation for couplings $V > J = 1$.  
	In contrast, the anti-correlation observed in $\rho_{1,3}$ is weaker and more susceptible to noise, while $\rho_{1,4}$ displays a positive correlation. 
	While stable triplet states have been described in literature~\cite{2010Valiente,2013Kornilovitch}, quadruplet states and larger clusters do not emerge as stable outcomes in the dynamic process.
	
	Given its prominence, we will focus on the pairing process from here on, rather than discussing general cluster formation.
	The underlying mechanism is the same as described in~\cite{2024Daumann}, where prethermalization was identified as a result of an effective attractive interaction caused by repulsive coupling between different particle species.
	From the perspective of the lighter particles, the presence of heavy particles manifests as potential barriers with height $V$.  
	When $V$ approaches the order of their kinetic energy $J$, the low-lying states of the light subsystem energetically favor clustered configurations of the heavy particles, minimizing momentum uncertainty.  
	This, in turn, exerts a Casimir-like force on the heavy particles, inducing a short-range attraction that ultimately leads to the observed pairing.
	
	The driving potential $\varepsilon_n$ can be determined using the Schrödinger equation for the light particles within the Born-Oppenheimer approximation~\cite{1927Born}, provided that the condition $J \gg J'$ holds:  
	\begin{equation}
		\mathcal{H}'\ket{\phi_n[\chi_m]}\otimes\ket{\chi_m} = \varepsilon_n[\chi_m]\ket{\phi_n[\chi_m]}\otimes\ket{\chi_m}\,.
		\label{eq:hamiltonianBO}
	\end{equation}
	In the reduced Hamiltonian $\mathcal{H}'$, the $J'$ terms from Eq.~(\ref{eq:hamiltonian}) are neglected, meaning that the heavy particles are treated as static, entering the equation as parametric configurations $\ket{\chi_m}$. 
	$\ket{\phi_n[\chi_m]}$ are the corresponding eigenstates of the light subsystem.
	The Born-Oppenheimer approximation assumes that the light subsystem has already relaxed to its ground state before the dynamics of the heavy subsystem come into play.  
	The heavy particles then move through the resulting energy landscape $\varepsilon_0[\chi_m]$, which can be used to estimate the stability of heavy-particle pairs.  
	The difference in energy $E_\text{gain}$, defined as the energy the system gains by forming pairs, is given by:  
	\begin{equation}
		E_\text{gain} = \varepsilon_0[\chi_\text{free}] - \varepsilon_0[\chi_\text{pair}] \
		\begin{cases}
			\ > 0 & : \text{stable}\\
			\ \leq 0 & : \text{unstable}
		\end{cases}
	\end{equation}
	Here, $\ket{\chi_\text{free}}$ denotes a configuration in which two heavy particles are at their maximum possible distance from each other, while $\ket{\chi_\text{pair}}$ corresponds to the paired state.  
	The stability depends on the coupling strength $V$ and is shown in Fig.~\ref{fig:pairing}f.  
	For couplings $V > J$, the energy gain becomes positive, making the formation of pairs energetically favorable.
	This aligns well with the observed perfect anti-correlations in Fig.~\ref{fig:pairing}e.
	
	\begin{figure}
		\includegraphics[width=1.0\linewidth]{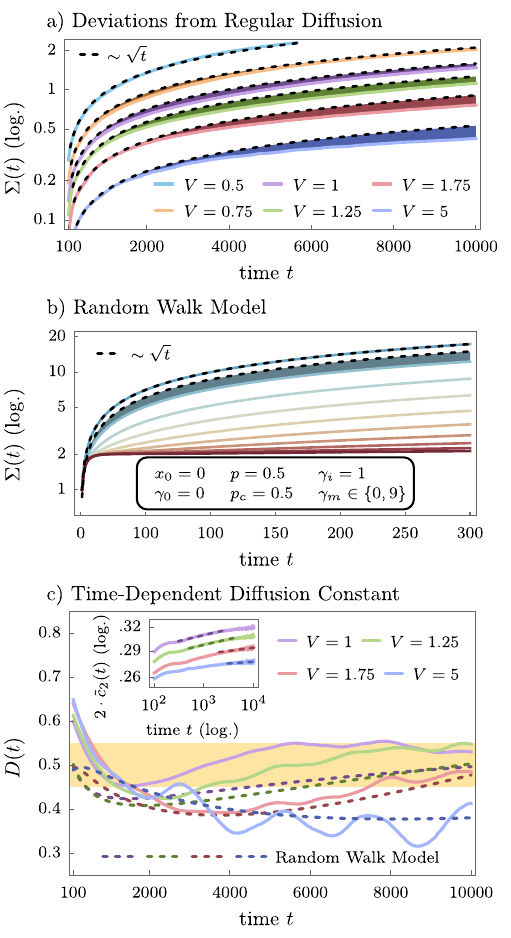}
		\caption{a) Square root of the mean squared displacement $\Sigma(t)$ for varying couplings $V$. Higher $V$ leads to more pronounced subdiffusion which persists for longer. Deviations from the expected diffusive behavior are emphasized as dark shaded areas. b) Classical random walk model with clustering mechanism for different maximum clusters $\gamma_m$. The deviation from $\sqrt{t}$-scaling is shown only for $\gamma_m = 1$ to maintain clarity. c) Comparison between time-dependent diffusion constants $D(t)$ in Hubbard (bright) and Random Walk Model (dark, dashed). The inset shows normalized pair data $\tilde{c}_2(t)$ for parameter determination. Dark dashed lines indicate the slopes for $\gamma_i$.}
		\label{fig:subdiffusion}
	\end{figure}
        
	{\it Subdiffusion} - As discussed in detail in~\cite{2024Daumann}, such a pairing process temporarily slows down or even halts the dynamics, as particles become trapped in $\varepsilon_0$ for timescales of up to $\sim J'^{-2}$, after which they may continue to delocalize as pairs in a second order process of $J'$.  
	
	Fig.~\ref{fig:subdiffusion}a shows time evolution in terms of the square roots of the mean squared displacement $\Sigma$ for single particle density peaks (setting (i)) for different $V$.
	It becomes apparent that the pairing regime for $V > J$ exhibits a distinct temporary subdiffusive stage at times $t \gg J'^{-1}$, during which $\Sigma$ increases more slowly than $\sqrt{t}$, the characteristic scaling expected for standard diffusion.
	With increasing coupling strength $V$, these deviations from the $\sqrt{t}$ behavior become more pronounced and persist for longer periods.  
	It is important to note that the observed subdiffusive stage does not persist until full thermalization but only until pairing, or more generally, until the cluster formation process is complete.
	From that point onward, transport resumes as a regular diffusive process - a behavior also observed in~\cite{2023Li}.
	
	The same characteristics arise when incorporating the identified pairing mechanism into a classical 1D random walk model.
	The procedure for this model is as follows:
	{\small
		\begin{itemize}[noitemsep,topsep=0pt]
			\item[0.)] Initialize: position $x_0\rightarrow x_t$, cluster size $\gamma_0\rightarrow \gamma_t$
			\item[1.)] A particle may hop left, right, or remain in place with probabilities $\{p^{1+\gamma_t},p^{1+\gamma_t},1-2p^{1+\gamma_t}\}$: $x_{t+1} = x_t+\{-1,+1,0\}$
			\item[2.)] A particle may form a cluster with probabilities $\{p_c,1-p_c\}$: $\gamma_{t+1}=\gamma_t+\gamma_i\cdot\{1,0\}$ is updated with increment $\gamma_i$ if $\gamma_{t+1}<\gamma_m$
			\item[3.)] Update $t\rightarrow t+1$ $\Rightarrow$ return to 1.)
		\end{itemize}} 
	The results for a classical system with integer cluster sizes $\gamma_t$ and an increment of $\gamma_i=1$ are shown in Fig.~\ref{fig:subdiffusion}b for different peak cluster sizes $\gamma_m$.
	$\gamma_m=0$ corresponds to a Brownian particle delocalizing steadily with $\sqrt{t}$.
	Allowing clustering by $\gamma_m>0$ leads to a decrease in hopping probability while $\gamma_t$ may increase.
	During this phase a slowed-down, subdiffusive transport up to the prethermalization stage can be observed.  
	Only when maximum $\gamma_m$ is reached does the transport become diffusive again at timescales $\sim p^{-(1+\gamma_m)}$.
	
	A qualitative comparison between behavior of diffusion constants in the Hubbard and random walk model is shown in Fig.~\ref{fig:subdiffusion}c.
	Most random walk model parameters can be extracted quite straightforwardly from the Hubbard model.
	The hopping probability is given by $p=J'=0.01$.
	Clustering probability $p_c$ is estimated by multiplying $p$ with the background particle density $n_\text{bg}$: $p_c=p\cdot n_\text{bg}=0.004159$.
	$n_\text{bg}$ is calculated in the $y\rightarrow\infty$ limit for one distinctly localized particle.
	$x_0$ is set to $0$ as before.
	
	Allowing initial cluster size $\gamma_0$, the increment $\gamma_i$ and the maximum cluster value $\gamma_m$ to be real quantities, the random walk model can take into account the quantum nature of the Hubbard model.
	These parameters are extracted from normalized cluster information $\tilde{c}_m(t)$ in the $y\rightarrow\infty$ limit.
 	In contrast to $c_m(t)$, the $\tilde{c}_m(t)$'s are normalized and fulfill $\sum_m\tilde{c}_m(t)=1\ \forall \ t$.
	All parameters are found in the pairing data $\tilde{c}_2(t)$, shown in the inset of Fig.~\ref{fig:subdiffusion}c.
	$\gamma_0$ and $\gamma_i$ vary with coupling $V$. 
	$\gamma_0$ can be calculated by the amount of paired particles $2\cdot \tilde{c}_2(t)$ at the timescale $t=J'^{-1}$ when heavy particle dynamics sets in.
	The slope of $2\cdot \tilde{c}_2(t)$ before saturation is used for $\gamma_i$.
	The resulting parameters $\gamma_0$ and $\gamma_i$ read as follows:
	{\small
	\begin{table}[h]
		\begin{tabular}{c|cccc}
			$V$ & $1.0$ & $1.25$ &$1.75$& $5.0$\\\hline 
			$\gamma_0$ & $0.2894$ & $0.2778$ & $0.2650$ & $0.2587$\\
			$\gamma_i$ & $0.0210$ & $0.0206$ & $0.0116$ & $0.0060$
		\end{tabular}
	\end{table}}\\
	In contrast, $\gamma_m$ is independent of specific $V$ values and approximated by the free particle limit $V\rightarrow0$ as $2\cdot \tilde{c}_2(t)$ for $t\rightarrow\infty$, yielding $\gamma_m=0.3927$.
	This approximation serves as a rough estimate since exact limits for $V\gg 0$ are inaccessible due to the required long time evolution.
	
	The random walk model reproduces the transient subdiffusive stage observed for $V>J$.
	The agreement is rather poor for weak couplings.
	However, with increasing $V$, where subdiffusive behavior becomes more pronounced, the agreement between the random walk and Hubbard model improves significantly.
	This comparison corroborates the hypothesis that particle pairing causes subdiffusion of heavy particles in the imbalanced Hubbard model.
        
	{\it Conclusion} - We showed that small clusters of heavy particles up to triplets form in the imbalanced Hubbard model when the coupling between light and heavy particles becomes strong. 
	The pairing process is predominant, and the stability of pairs can be understood within the Born-Oppenheimer approximation.
	This clustering mechanism induces subdiffusive transport, which could be quantitatively reproduced within an random walk model parameterized by quantum mechanical data from the Hubbard model.
    Thus, our work provides evidence for a mechanism of subdiffusion in the imbalanced Hubbard model, which is unrelated to the Griffiths picture.
	
	We thank Christian Eidecker-Dunkel, Patrick Vorndamme, Robin Steinigeweg and Roderich Moessner for valuable discussions. Financial support from the DFG via the research group FOR2692, grant number
	397171440 is gratefully acknowledged.
	 
	The data that support the findings of this article (Fig.~\ref{fig:pairing} and Fig.~\ref{fig:subdiffusion}) are openly available \cite{data}.

	\bibliography{ppcshpihm_bib}
	
\end{document}